\documentclass[10pt,letterpaper]{article}
\usepackage{opex3}
\usepackage{citesort}

\begin{document}

\title{Direct Measurement of Cyclotron Coherence Times of High-Mobility Two-Dimensional Electron Gases}

\author{X.~Wang$^1$, D.~J.~Hilton$^2$, J.~L.~Reno$^3$, D.~M.~Mittleman$^1$, and J.~Kono$^{1,*}$}
\address{$^1$Department of Electrical and Computer Engineering, Rice University, \\ Houston, Texas 77005, USA
\\ $^2$Department of Physics, University of Alabama at Birmingham, Alabama 35294, USA
\\$^3$Sandia National Laboratories, Albuquerque, New Mexico 87185, USA}

\email{kono@rice.edu} 



\begin{abstract}
We have observed long-lived ($\sim$30~ps) coherent oscillations of charge carriers due to cyclotron resonance (CR) in high-mobility two-dimensional electrons in GaAs in perpendicular magnetic fields using time-domain terahertz spectroscopy.  The observed coherent oscillations were fitted well by sinusoids with exponentially-decaying amplitudes, through which we were able to provide direct and precise measures for the decay times and oscillation frequencies simultaneously.  This method thus overcomes the CR saturation effect, which is known to prevent determination of true CR linewidths in high-mobility electron systems using Fourier-transform infrared spectroscopy.
\end{abstract}


\bibliographystyle{osajnl} 

%
%

\section{Introduction}
Cyclotron resonance (CR) in two-dimensional electron gases (2DEGs) formed in semiconductor inversion layers, heterojunctions, and quantum wells has been studied for many years, exhibiting a number of surprising phenomena~\cite{PetrouMcCombe91LLS,Nicholas94HOS,Kono01MMR,KonoMiura06HMF}.  These phenomena (so-called ``CR anomalies''), observed as unexpected line broadening, splittings, and shifts, are believed to arise from the interplay of band structure, disorder, and interaction effects.  As the sample mobility becomes higher, CR lines are expected to become narrower, which should permit more systematic studies to precisely determine the origins of these effects.  CR has indeed been studied in a variety of 2DEG systems with a wide range of mobilities (10$^{4}$ to 10$^{6}$ cm$^{2}$/Vs).  However, CR studies of {\em high-mobility ($>$10$^{6}$~cm$^{2}$/Vs)} 2DEGs using traditional Fourier-transform infrared spectroscopy have had limited success due, at least partially, to the CR saturation effect~\cite{ChouetAl88PRB}.  In the high mobility/high carrier density limit, the 2DEGs behave as a metallic mirror, reflecting off most of the incident light at the CR peak, and thus, only a small amount of the incident light is transmitted by the high mobility 2DEG~\cite{StudenikinetAl05PRB,Mikhailov04PRB}.  This effect results in an undesirable broadening of transmittance linewidths, relative to the true CR linewidths, as shown in Section~\ref{theory} below.

Here, we demonstrate how the CR saturation effect can be overcome using terahertz (THz) time-domain techniques.  First, through theoretical simulations while systematically increasing the mobility, we show that, in the high mobility limit, the calculated power transmittance peak is broader than the CR peak in the calculated real part of the complex conductivity by an amount dependent on the carrier density.  However, the linewidth determined from the conductivity keeps decreasing with increasing mobility, always representing the true CR linewidth.  In our experiments, described in Section~\ref{results}, we observed long-lived time-domain CR oscillations in a high-mobility (3.7 $\times$ 10$^6$~cm$^2$Vs) GaAs/AlGaAs 2DEG, measured with our time-domain THz magneto-spectroscopy system~\cite{WangetAl07OL,WangetAl10NP}.  Quantum mechanically, these time-domain oscillations are due to a resonant optical transition across the Fermi energy between the highest filled Landau level and the lowest unfilled Landau level.  The observed coherent oscillations were fitted well by sinusoids with exponentially-decaying amplitudes, through which we were able to directly determine the CR decay times and CR frequencies simultaneously, thereby overcoming the CR saturation effect.

\section{Cyclotron resonance saturation effect: theoretical simulations}
\label{theory}

\begin{figure}
\centering
\includegraphics[width=11cm]{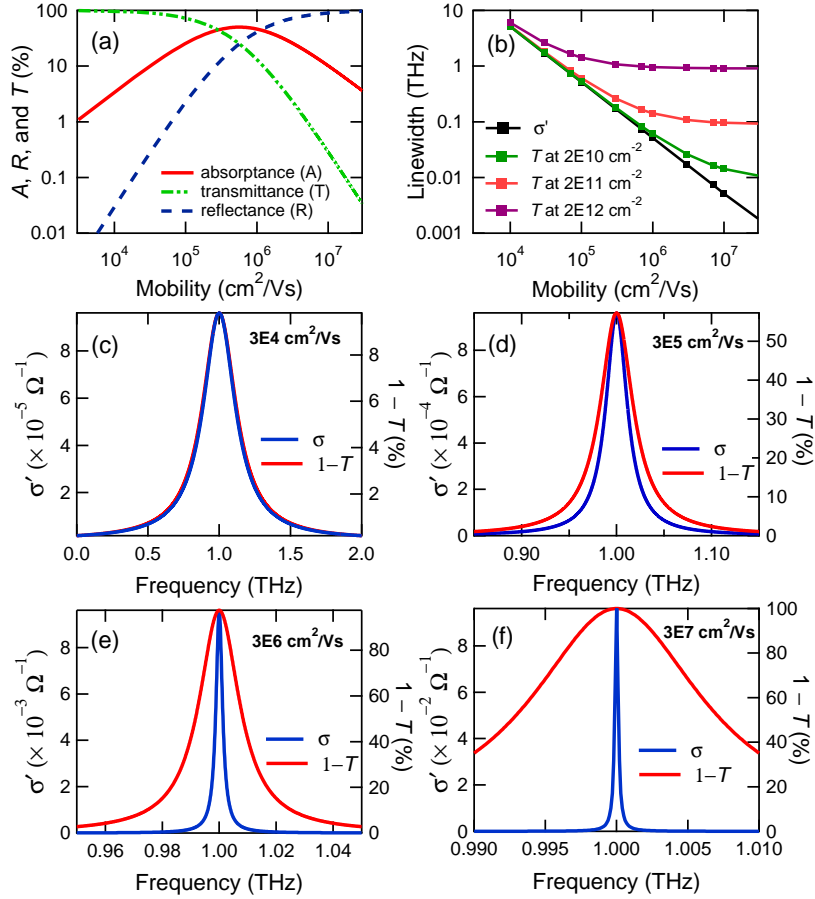}
\centering \caption{\textbf{(a)} Calculated absorptance $A$, reflectance $R$, and transmittance $T$ versus CR mobility at a fixed density $n$ = 2 $\times$ 10$^{11}$ cm$^{-2}$.  \textbf{(b)} Calculated linewidths of the real part of the conductivity and transmittance at $n$ = 2 $\times$ 10$^{10}$, 2 $\times$ 10$^{11}$, and 2 $\times$ 10$^{12}$ cm$^{-2}$.  The mobility was changed from 1 $\times$ 10$^{4}$ to 1 $\times $10$^{8}$ cm$^{2}$/Vs in \textbf{(a)} and \textbf{(b)}.  \textbf{(c, d, e, and f)} Calculated real part of the conductivity $\sigma'$ (blue lines) and $1 - T$ (red lines) for different mobilities: \textbf{(c)} 3 $\times$ 10$^{4}$, \textbf{(d)} 3 $\times$ 10$^{5}$, \textbf{(e)} 3 $\times$ 10$^{6}$, and \textbf{(f)} 3 $\times$ 10$^{7}$ cm$^{2}$/Vs, respectively.  The deviation of the transmittance peak linewidth from the real conductivity linewidth becomes more significant with increasing mobility. }
\label{fig1}
\end{figure}

We approximate the sample as a 2DEG layer between two GaAs substrate layers, which have an admittance, $Y = n_{{\rm_{GaAs}}}/Z_{0}$, where $Z_{0}$ = 377~$\Omega$ is the impedance of free space and $n_{\rm_{GaAs}}$ = 3.60 is the refractive index of GaAs at THz frequencies~\cite{GrischkowskyetAl90JOSAB}.  When the wavelength and skin depth
are much larger than the 2DEG thickness, the THz electric field can be considered to be uniform.  Namely, the THz electric field is not a function of 2DEG thickness, and the 2DEG can be treated as an interface between the two substrates.  We employ the thin conducting film approximation described in Ref.~\cite{Nussbook}, in which the transmission coefficient, $T'$, and reflection coefficient, $R'$, for a left-circularly polarized (or cyclotron active) THz field are given by
\begin{equation}\label{eq:T}
T' = \frac{E\bigl(B\bigr)}{E\bigl(0\bigr)} = \frac{2Y}{2Y + \sigma}
\end{equation}
\begin{equation}\label{eq:R}
R' = \frac{\sigma}{2Y + \sigma}
\end{equation}
where $E(B)$ is the transmitted THz field in an applied external magnetic field $B$ and $E(0)$ is the transmitted THz field at 0~T.  Therefore, the transmittance $T$ is given by $T = |T'|^2$, and the reflectance $R$ is given by $R = |R'|^2$.  The absorptance $A$ can then be calculated through $A = 1 - T - R$.  Within the Drude model, the complex
conductivity, $\sigma$, of the 2DEG layer in magnetic fields is given by:
\begin{equation}\label{eq:sigma}
\sigma =\sigma'+ i\sigma''=
\frac{ne^2}{m^*}\frac{1/\tau_{\rm CR}}{1/\tau_{\rm CR}^2+4\pi^2(f-f_{\rm c})^2} + i\frac{ne^2}{m^*}\frac{2\pi(f-f_{\rm c})}{1/\tau_{\rm CR}^2+4\pi^2(f-f_{\rm c})^2}
\end{equation}
where $n$ is the electron sheet density in the 2DEG, $m^*$ is the electron effective mass, $\tau_{\rm CR}$ is the CR decay time, and $f_c$ is the CR frequency.  The CR mobility is defined as $\mu_{\rm CR} = e\tau_{\rm CR}/m^*$.  We only consider the left circularly polarized (or cyclotron active) component because the right circularly polarized
component does not contribute to cyclotron resonance.  

Absorptance ($A$), transmittance ($T$), and reflectance ($R$) at the CR frequency $f_c$ were thus calculated with varying mobilities from 3 $\times$ 10$^4$ to 3 $\times$ 10$^7$ cm$^2$/Vs at a fixed density of $n$ = 2 $\times$ 10$^{11}$~cm$^{-2}$.  The results are shown in Fig.~\ref{fig1}(a).  At low mobilities, the absorptance $A$ increases with the mobility followed by a saturation at $\mu$ $\sim$ 5.66 $\times$ 10$^5$~cm$^2$/Vs and then decreases with the further increase of the mobility.  The reflectance $R$ shows a monotonic increase while the transmittance $T$ decreases with increasing mobility.  For a mobility of 5 $\times$ 10$^6$~cm$^2$/Vs, the absorption at the resonance is $\sim$18.28\%, the transmittance is $\sim$1.04\%, and the reflectance is $\sim$80.68\%.  It is clear that most of the
incident power is reflected by the 2DEG and only a small amount is transmitted in this regime, where the mobility is sufficiently high.  A low transmittance and high reflectance in high-mobility samples make Fourier-transform infrared techniques difficult and points to the need for an experimental technique with a sufficiently high signal-to-noise ratio to isolate the small transmission signal.  Time-domain THz spectroscopy is an ideal tool for measuring CR in these high-mobility samples because of its high sensitivity.

Furthermore, the CR saturation effect affects the determination of the true CR linewidth from transmittance spectra. Figure~\ref{fig1}(b) shows the calculated linewidths both in transmittance and conductivity spectra plotted versus the mobility for several carrier densities ($n$ = 2 $\times$ 10$^{10}$, 2 $\times$ 10$^{11}$, and 2 $\times$ 10$^{12}$ cm$^{-2}$).  It can be seen from Equation~\ref{eq:sigma} that the linewidth determined from the conductivity peak is inversely proportional to the CR decay time.  In the low density and low mobility limit, a direct relation is present between the transmittance linewidth and the conductivity linewidth; in this limit, the true conductivity linewidth can be determined from the transmittance linewidth.  However, with increasing mobility, the transmittance linewidth starts deviating from the true CR linewidth, and eventually saturates.  The higher the sample density, the larger the deviation.  Figures~\ref{fig1}(c)-(f) show the real part of the conductivity ($\sigma'$) and $1-T$ versus frequency for four different mobilities at a fixed density of 2 $\times$ 10$^{11}$~cm$^{-2}$.  Again, the deviation is seen to increase with increasing mobility.  It is clear from Equation~\ref{eq:sigma} that the CR linewidth can be accurately determined by examining the linewidth of the conductivity peak or the lifetime of time-domain oscillations.  Hence, in the high mobility and/or high carrier density limit, the advantage of time-domain THz spectroscopy results from its ability to determine the full complex conductivity, $\sigma$ in the frequency domain, and/or a direct measurement of the free induction decay of CR in the time domain as shown in the following section.

\begin{figure}
\centering
\includegraphics[width=10cm]{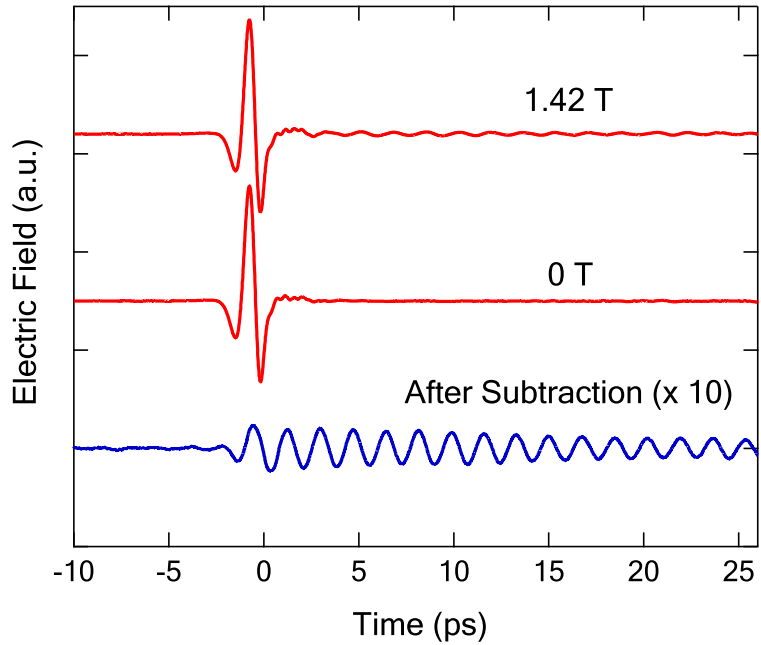}
\centering \caption{THz waveforms transmitted through a high-mobility 2DEG at 0~T and at 1.42~T at 1.6~K.  The coherent cyclotron resonance oscillations, present only in the 1.42~T waveform, are isolated by subtracting the 0~T waveform from the 1.42~T waveform.} \label{fig2}
\end{figure}

\begin{figure}
\centering
\includegraphics[width=10cm]{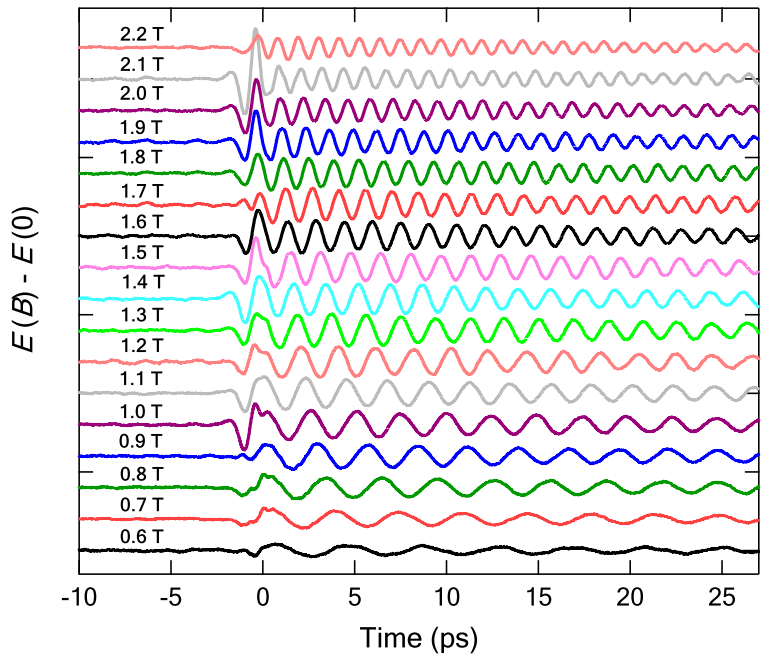}
\centering \caption{Coherent cyclotron resonance oscillations of a GaAs 2DEG at magnetic fields from 0.6 to 2.2~T at 1.6~K, measured by time-domain THz magneto-spectroscopy.  The traces are vertically offset for clarity.  These CR oscillations are isolated by subtracting the transmitted THz waveform at 0~T from the transmitted THz waveforms at different magnetic fields.} \label{fig3}
\end{figure}

\section{Experimental methods}
\label{methods}
In order to experimentally demonstrate the utility of time-domain THz spectroscopy to determine the true CR linewidth of a high-mobility 2DEG and overcome the saturation effect, we used a time-domain THz magneto-spectroscopy system~\cite{WangetAl07OL,WangetAl10NP}.  The system consisted of a chirped pulse amplifier (CPA-2001, Clark-MXR, Inc.) with a wavelength of 800~nm and a pulse width of $\sim$200~fs and a pair of $<$110$>$ ZnTe crystals to generate and detect coherent radiation from 0.1 to 2.6~THz through surface rectification and electro-optic sampling, respectively.  A shaker, operating at 2~Hz, provided time delays up to 80~ps.  We averaged over 1000 scans for acquiring each time-domain waveform.  We used a magneto-optical cryostat to generate magnetic fields up to 10~T, and the sample temperature was kept at 1.6~K for this study.  We performed our experiments in a Faraday geometry with both the incident THz wave and the external magnetic field perpendicular to the sample surface.  The sample was a GaAs 2DEG with $n$ = 2 $\times$ 10$^{11}$~cm$^{-2}$ and $\mu$ = 3.7 $\times$ 10$^6$~cm$^2$V$^{-1}$s$^{-1}$ at 4 K, which was physically large enough (0.6 $\times$ 10 $\times$ 15 mm$^3$) to neglect any finite-size-induced magneto-plasmonic shift of CR frequencies (see, e.g., \cite{PetrouMcCombe91LLS,Nicholas94HOS,Kono01MMR}).


\section{Experimental results and discussion}
\label{results}
We measured transmitted THz electric fields through the 2DEG sample in magnetic field from 0 to 2.2~T at~1.6~K.  In order to deduce the magnetic-field-induced changes to the transmitted waveforms, we subtracted the 0~T waveform from
each finite-field waveform.  One example is shown in Fig.~\ref{fig2}, showing clear oscillations, after subtracting the 0~T waveform from the 1.42~T waveform.  The magnetic-field-induced oscillations, extracted in this manner, from 0.6~T to 2.2~T are plotted in Fig.~\ref{fig3}.  The frequency of the oscillations is clearly seen to increase with increasing magnetic field.  These oscillations result from cyclotron resonance, i.e., an optical transition between the highest-filled Landau level and the lowest-unfilled Landau level.

We fit these CR oscillations with an exponentially-decaying sinusoid:
\begin{equation}\label{eq:TDFit}
E(t) = E_0 + A\exp(-t/\tau_{\rm CR})\sin(2{\pi}f_{\rm c}+\theta_{0})
\end{equation}
where $\tau_{\rm CR}$ is the CR decay time, $f_{\rm c}$ is the cyclotron frequency, $E_0$ is a DC offset, $A$ is
the initial field amplitude, and $\theta_0$ is the initial phase of the oscillations.  We fitted the CR oscillations within a 26~ps time window, because of two multiple reflection peaks that appear at $\sim$28.3~ps and $\sim$44.2~ps, respectively, due to magnet windows.  They induce additional coherent oscillations that overlap those induced by the main THz pulse.  In addition, the oscillations from $-$2 to 2 ps were not used for fitting because they contain extra information about initial absorption of the incident THz pulse.  The 26~ps window chosen contains enough periods for fitting accuracy.

\begin{figure}
\centering
\includegraphics[width=10cm]{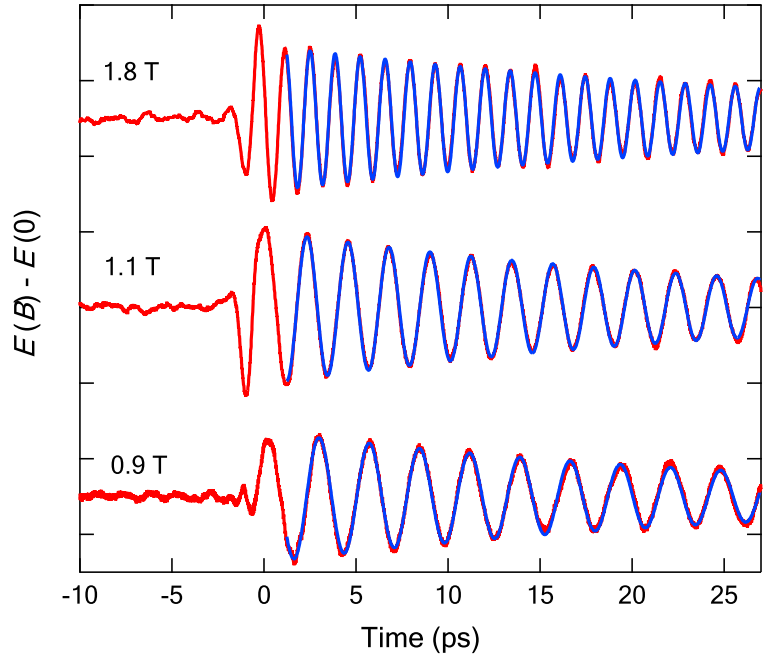}
\centering \caption{Time-domain cyclotron resonance oscillations at 1.8, 1.1, and 0.9~T obtained by subtracting the transmitted THz waveform at 0~T from the transmitted THz waveforms at these three magnetic fields (red curves).  We fit these oscillations using an exponentially-decaying sinusoid (see Equation~\ref{eq:TDFit}).  Fitting curves are shown as blue curves.} \label{fig4}
\end{figure}

Several fitting examples are shown in Fig.~\ref{fig4}.  Here, the data recorded at magnetic fields of 1.8~T, 1.1~T, and 0.9~T were fitted with Equation~\ref{eq:TDFit}, showing excellent agreement with the experimental data.  The obtained fitting parameters are summarized in Table~\ref{table:table1} with estimated uncertainties.  One can see that standard deviations are less than 1~$\%$ for CR decay times ($\tau_{CR}$), while the CR frequency ($f_{c}$) is determined with much smaller uncertainties ($<$0.01\%) as expected.  The extracted cyclotron frequency $f_{c}$ vs.~magnetic field shows a linear relationship (not shown), also as expected from the relation $2\pi f_{\rm c} = eB/m^*$.  From the slope of the linear curve, we obtained an effective mass of $0.068m_0$, which is the standard value for electrons in a GaAs 2DEG.  Also included in \ref{table:table1} are the CR decay times that would be obtained from transmittance linewidths (as fully discussed in Section~\ref{theory}.  Clear differences between the CR decay times obtained through time-domain CR oscillations and power transmittance are seen.

\begin{table}[ht]
\caption{CR decay times ($\tau_{\rm CR}$) and CR frequencies ($f_{\rm c}$) obtained through fitting time-domain CR oscillations at 1.8, 1.1, and
0.9~T.  Standard deviations of the fittings are also shown.  For comparison, CR decay times that would be obtained 
from transmittance linewithds ($\tau_{\rm Tr}$) are shown in the last column.} 
\centering 
\begin{tabular}{c c c c c} 
\hline\hline 
Magnetic Field (T) & $\tau_{\rm CR}$ (ps) & $f_{\rm c}$ (GHz) & $\tau_{\rm Tr}$ (ps)\\ [0.5ex]  
\hline 
1.8 & 30.8 $\pm$ 0.2 & 736.29 $\pm$ 0.04 & 12.9 \\ 
1.1 & 28.3 $\pm$ 0.1 & 449.33 $\pm$ 0.03 & 12.4 \\
0.9 & 27.4 $\pm$ 0.2 & 366.74 $\pm$ 0.05 & 12.3 \\[1ex] 
\hline 
\end{tabular}
\label{table:table1} 
\end{table}

\section{Conclusions}
We have observed time-domain cyclotron resonance oscillations in a GaAs/AlGaAs 2DEG using a THz magneto-spectroscopy system from 0.6 to 2.2~T at 1.6 K. By fitting these oscillations with an exponentially-decaying sinusoid, we were able to directly determine the cyclotron resonance decay time and frequency.  We demonstrated, both experimentally and theoretically, that using time-domain THz magneto-spectroscopy one can overcome the cyclotron resonance saturation effect and determine the true CR linewidths in high-mobility 2DEG samples.

\section{Acknowledgments}
This work was supported by the National Science Foundation through Award Nos.~DMR-0134058, DMR-0325474, and OISE-0530220 and the Robert A. Welch Foundation through Grant No.~C-1509.  Sandia is a multiprogram laboratory
operated by Sandia Corporation, a Lockheed Martin Company, for the National Nuclear Security Administration
under contract DE-AC04-94AL85000.


\begin{thebibliography}{10}
\newcommand{\enquote}[1]{``#1''}

\bibitem{PetrouMcCombe91LLS}
A.~Petrou and B.~D. McCombe, \enquote{Magnetospectroscopy of confined
  semiconductor systems,} in \enquote{Landau Level Spectroscopy, Volume 27.2 of
  Modern Problems in Condensed Matter Sciences,} , G.~Landwehr and E.~I.
  Rashba, eds. (Elsevier Science, Amsterdam, 1991), pp. 679--775.

\bibitem{Nicholas94HOS}
R.~J. Nicholas, \enquote{Intraband optical properties of low-dimensional
  semiconductor systems,} in \enquote{Handbook on Semiconductors, Vol.~2
  ``Optical Properties'',} , M.~Balkanski, ed. (Elsevier, Amsterdam, 1994), pp.
  385--461.

\bibitem{Kono01MMR}
J.~Kono, \enquote{Cyclotron resonance,} in \enquote{Methods in Materials
  Research,} , E.~N. Kaufmann, R.~Abbaschian, A.~Bocarsly, C.-L. Chien,
  D.~Dollimore, B.~Doyle, A.~Goldman, R.~Gronsky, S.~Pearton, and J.~Sanchez,
  eds. (John Wiley \& Sons, New York, 2001), chap. 9b.2.

\bibitem{KonoMiura06HMF}
J.~Kono and N.~Miura, \enquote{Cyclotron resonance in high magnetic fields,} in
  \enquote{High Magnetic Fields: Science and Technology, Volume III,} ,
  N.~Miura and F.~Herlach, eds. (World Scientific, Singapore, 2006), pp.
  61--90.

\bibitem{ChouetAl88PRB}
M.~J. Chou and D.~C. Tsui, \enquote{Cyclotron resonance of high-mobility
  electrons at extremely low densities,} Phys. Rev. B \textbf{37}, 848--854
  (1988).

\bibitem{StudenikinetAl05PRB}
S.~A. Studenikin, M.~Potemski, A.~Sachrajda, M.~Hillke, L.~N. Pfeiffer, and
  K.~W. West, \enquote{Microwave-induced resistance oscillations on a
  high-mobility two-dimentional electron gas: exact waveform,
  absorption/reflection and temperature damping,} Phys. Rev. B \textbf{71},
  245313--245322 (2005).

\bibitem{Mikhailov04PRB}
S.~A. Mikhailov, \enquote{Microwave-induced magnetotransport phenomena in
  two-dimentional electron systmes: important of electodynamic effects,} Phys.
  Rev. B \textbf{70}, 165311--165315 (2004).

\bibitem{WangetAl07OL}
X.~Wang, D.~J. Hilton, L.~Ren, D.~M. Mittleman, J.~Kono, and J.~L. Reno,
  \enquote{Terahertz time-domain magnetospectroscopy of a high-mobility
  two-dimensional electron gas,} Optics Lett. \textbf{32}, 1845--1847 (2007).

\bibitem{WangetAl10NP}
X.~Wang, A.~A. Belyanin, S.~A. Crooker, D.~M. Mittleman, and J.~Kono,
  \enquote{Interference-induced terahertz transparency in a semiconductor
  magneto-plasma,} Nature Physics \textbf{6}, 126--130 (2010).

\bibitem{GrischkowskyetAl90JOSAB}
D.~Grischkowsky, S.~Keiding, M.~V. Exter, and C.~Fattinger,
  \enquote{Far-infrared time-domain sepctroscopy wiht terahertz beams of
  dielectrics and semiconductors,} J. Opt. Soc. Am. B \textbf{7}, 2006--2015
  (1990).

\bibitem{Nussbook}
M.~C. Nuss and J.~Orenstein, \enquote{Terahertz time-domain spectroscopy,} in
  \enquote{Millimeter and Submillimeter Wave Spectroscopy of Solids,}
  (Springer-Verlag, Berlin, 1998), pp. 7--50.

\end{thebibliography}
\end{document}